\documentclass[runningheads]{llncs}
\usepackage[T1]{fontenc}
\usepackage{graphicx}
\usepackage{cite}
\usepackage{comment}
\usepackage{amsmath,amssymb,amsfonts}
\usepackage{algorithmic}
\usepackage{textcomp}
\usepackage{xcolor}
\usepackage{subcaption}
\usepackage{listings}
\usepackage{url}
\usepackage{hyperref}
\usepackage{ulem}
\usepackage{todonotes}
\usepackage{float}

\newcommand{\orcid}[1]{\href{https://orcid.org/#1}{\includegraphics[height=10pt]{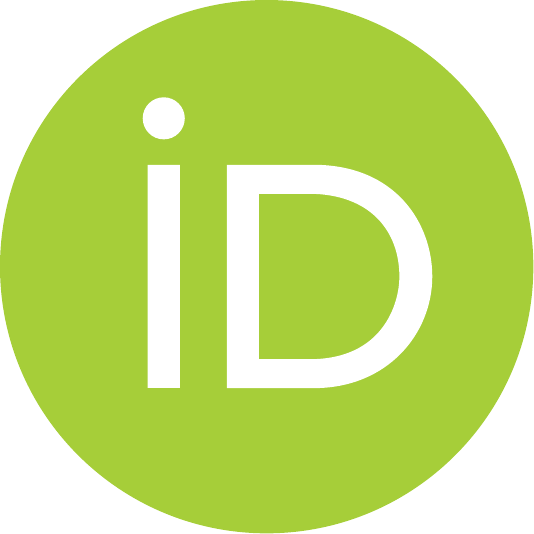}}}

\def\BibTeX{{\rm B\kern-.05em{\sc i\kern-.025em b}\kern-.08em
    T\kern-.1667em\lower.7ex\hbox{E}\kern-.125emX}}

\begin{document}

\title{An Initial Evaluation of Distributed Graph Algorithms using NWGraph and HPX}

\author{Karame Mohammadiporshokooh \orcidID{0009-0000-8349-3389}\and
Panagiotis Syskakis \and
Hartmut Kaiser\orcidID{0000-0002-8712-2806}}

\titlerunning{Distributed Graph Algorithms with NWGraph and HPX}
\authorrunning{K. Mohammadiporshokooh et al.}

\institute {Louisiana State University, Baton Rouge, LA, 70810, USA}

\maketitle
\begin{abstract}
Graphs are central to modeling relationships in scientific computing, data analysis, and AI/ML, but their growing scale can exceed the memory and compute capacity of single nodes, requiring distributed solutions. Existing distributed graph frameworks, however, face fundamental challenges: graph algorithms are latency-bound, suffer from irregular memory access, and often impose synchronization costs that limit scalability and efficiency. In this work, we present a distributed implementation of the NWgraph library integrated with the HPX runtime system. By leveraging HPX’s asynchronous many-task model, our approach aims to reduce synchronization overhead, improve load balance, and provide a foundation for distributed graph analytics. We evaluate this approach using two representative algorithms: Breadth-First Search (BFS) and PageRank. Our initial results show that BFS achieves better performance than the distributed Boost Graph Library (BGL), while PageRank remains more challenging, with current implementations not yet outperforming BGL. These findings highlight both the promise and the open challenges of applying asynchronous task-based runtimes to graph processing, and point to opportunities for future optimizations and extensions.

\end{abstract}


\section{Introduction}

Graphs are widely used to represent relationships in domains such as scientific computing, social networks, and machine learning. As graph datasets grow in scale and complexity, they often exceed the capacity of single machines, motivating the need for distributed solutions. Unlike dense or sparse linear algebra problems, which are typically compute-bound or bandwidth-bound, graph algorithms are largely latency-bound. Their irregular structure, fine-grained data dependencies, and unpredictable memory access patterns result in high communication and synchronization costs in distributed settings.

A number of distributed frameworks have been developed to address these challenges, including Pregel ~\cite{10.5555/2387880.2387883}, GraphX, AM++ ~\cite{7851538}, and the Boost Graph Library (BGL). While these systems provide abstractions for large-scale graph analytics, they continue to face issues of load imbalance, communication overhead, and synchronization bottlenecks. These limitations motivate exploration of alternative execution models that can better exploit fine-grained parallelism.

In this work, we investigate the integration of the NWgraph C++ library with the HPX runtime system. NWgraph provides a flexible foundation for graph representation and algorithm development, while HPX offers an asynchronous, task-based execution model designed to overlap computation and communication. Together, they provide a promising path toward improving scalability and efficiency in distributed graph processing.

To demonstrate this potential, we implemented distributed versions of two representative graph algorithms: Breadth-First Search (BFS), a traversal algorithm, and PageRank, a centrality algorithm. These algorithms capture distinct computational characteristics: BFS with level-synchronous frontier expansion, and PageRank with iterative, communication-intensive score propagation. We compare our HPX-based implementations against distributed BGL and present initial results. Our experiments show that BFS outperforms BGL on irregular graphs, while PageRank remains more challenging, with current results not yet surpassing BGL. These findings highlight both the promise and the open challenges of applying asynchronous runtimes to distributed graph analytics.

\section{Related work and Challenges}

Distributed graph processing frameworks aim to handle large graphs by partitioning the data across multiple machines and coordinating computation across them. Their design usually balances three factors: communication efficiency, workload distribution, and synchronization overhead.

One common approach is the Bulk Synchronous Parallel (BSP) model, as seen in Pregel\cite{10.1145/1807167.1807184} and GraphX. BSP provides a clean programming interface but requires global barriers at each superstep. This introduces synchronization delays and often amplifies load imbalance when graphs have skewed degree distributions. To address these issues, frameworks such as AM++\cite{7851538} introduce active messages that allow computation and communication to overlap. While this reduces barrier costs, it introduces new complexity in message handling, memory usage, and scheduling.

The Parallel Boost Graph Library (PBGL)\cite{DBLP:books/crc/22/EdmondsL22} and its successor PBGL 2.0\cite{edmonds2022parallel} extend this message-driven design to implement fundamental graph algorithms like BFS and shortest paths. PBGL shows how asynchronous execution can outperform BSP approaches, but it also highlights the difficulty of managing fine-grained communication at scale.

Despite these advances, distributed graph systems continue to face persistent challenges. Load imbalance arises when some nodes handle disproportionately large subgraphs or high-degree vertices. Communication overhead results from the irregular and data-driven nature of graph algorithms, which produce fine-grained, unpredictable message traffic. Synchronization costs remain a bottleneck in BSP-style systems, while fully asynchronous systems struggle with memory management and runtime overhead. At larger scales, fault tolerance is also an increasing concern~\cite{lumsdaine_challenges_2007, meng2024surveydistributedgraphalgorithms}.

Beyond CPU-based systems, GPU-focused frameworks such as Gunrock and Lux have demonstrated strong performance on single- or multi-GPU platforms. However, scaling these approaches across distributed environments remains difficult due to communication overhead and irregular frontier growth. Since our focus is on CPU-based distributed execution, we do not pursue GPU-based approaches further in this work.

Overall, existing distributed graph frameworks provide useful abstractions but leave open issues in scalability, load balance, and communication efficiency. These challenges motivate the need to explore new runtime systems and execution models for distributed graph processing.

\section{Framework Design and Implementation}

\subsection{NWGraph}
\label{sec:nwgraph}

NWGraph~\cite{Lumsdaine2021NWGraphAL,nwgraph_github} is a modern C++ library that provides a generic foundation for graph algorithms. Its design relies on C++20 concepts to define minimal type requirements for algorithms and data structures. This enables algorithms to be expressed in a flexible way that is not tied to a specific graph representation. For example, a graph can be described as a range of ranges: an outer range of vertices, where each vertex is associated with an inner range of neighbors. This model provides a clean and lightweight interface while supporting common operations such as adjacency queries and edge traversal.

The library also introduces a collection of traversal utilities, such as BFS and DFS ranges, which make it easier to express graph algorithms in terms of high-level views rather than custom iterators or visitor objects. In addition, NWGraph integrates with Intel’s Threading Building Blocks (TBB) to enable parallel execution within a single node. However, its parallelism is primarily limited to shared-memory systems. This limitation motivates the need to combine NWGraph with a distributed runtime to scale beyond a single machine.

\subsection{HPX Runtime System}
\label{sec:hpx}

HPX~\cite{hpx_joss_paper,hpx_stellar_group,kaiser_2024_598202} is a C++ standards-conforming runtime system for parallel and distributed computing. It follows the asynchronous many-task (AMT) model, where computations are decomposed into lightweight tasks (HPX-threads) that can be scheduled dynamically. ~\cite{wu2022quantifying, mohammadiporshokooh2025adaptively} HPX supports work stealing, global synchronization through futures, and an active global address space (AGAS) that enables transparent access to distributed objects.

For distributed graph algorithms, these features are particularly valuable. The asynchronous task model helps overlap communication with computation, reducing latency. AGAS provides a uniform way to distribute graph data across nodes while still allowing fine-grained access. HPX’s distributed containers and messaging system allow graph partitions to be processed locally while exchanging updates with minimal synchronization.

\section{Algorithms and Implementation}
\subsection{Breadth-First Search (BFS)}

Breadth-First Search (BFS) is one of the most widely used traversal algorithms and serves as a baseline for evaluating distributed graph frameworks. The algorithm explores the graph level by level, starting from a given source vertex. At each step, all vertices in the current “frontier” are expanded to discover their unvisited neighbors, which form the next frontier.

In distributed settings, BFS must manage graph partitions across multiple nodes. A common challenge is that frontier sizes change dramatically during the traversal, creating irregular workloads ~\cite{doi:10.1142/S0129626407002843}. For example, an early frontier may be very small, while later steps may involve millions of vertices. This variation makes load balancing difficult. Another challenge is that frontier expansion requires frequent communication to check and update visited vertices, which introduces high communication overhead. Prior studies have shown that synchronization costs further slow down distributed BFS~\cite{lumsdaine_challenges_2007, meng2024surveydistributedgraphalgorithms}.

\begin{figure}[h]
    \centering
\begin{lstlisting}[
    caption={Naïve Generic \textbf{Sequential} Breadth-First Search (BFS)},
    label={fig:bfs-1}
    ]
template <adjacency_list_graph Graph>
auto bfs(const Graph& G, id_t root) {
  std::deque<id_t> frontier{root};
  std::vector<id_t> parents(num_vertices(G), -1);
  parents[root] = root;
  while (!frontier.empty()) {
    std::deque<id_t> next;
    for (auto u : frontier) {
      for (auto e : G[u]) {
        id_t v = target(G, e);
        if (parents[v] == -1) {
          parents[v] = u;
          next.push_back(v);
        }
      }
    }
    frontier.swap(next);
  }
  return parents;
}
\end{lstlisting}
\end{figure}


\begin{figure}[H]
\centering
\begin{lstlisting}[
    caption={Naïve Generic \textbf{Distributed} Breadth First Search},
    label={fig:bfs-2}
    ]
template <typename Graph>
static void bfs_2(
    Graph G,
    hpx::partitioned_vector<id_t> parents,
    id_t source,
    id_t parent,
    size_t level)
{
    std::uint32_t here = hpx::get_locality_id();
    std::deque<std::tuple<id_t,size_t>> q1, q2;
    std::vector<hpx::future<void>> ops;

    // If not traversed before, add to queue
    if (set_parent(parents, source, parent, level))
    {
        q1.emplace_back(source, level);
    }

    while (!q1.empty()) {
        for (auto const& item : q1) {
            id_t u = std::get<0>(item);
            size_t lvl = std::get<1>(item);

            for (auto const& e : G[u]) {
                id_t v = target(G, e);
                auto dst = vertex_locality_id(G, v);

                if (dst == here) {
                    if (set_parent(parents, v, u, lvl))
                        q2.emplace_back(v, lvl + 1);
                } else {
                    ops.push_back(hpx::async(
                        bfs_2,
                        dst,
                        G,
                        parents,
                        v, u, lvl + 1          
                    ));
                }
            }
        }
        q1.swap(q2);
        q2.clear();
    }

    hpx::wait_all(ops);
}
    \end{lstlisting}
\end{figure}

HPX provides a distributed data structure, \texttt{hpx::partitioned\_vector}, which partitions an array across multiple localities in a distributed application. Its interface closely mirrors that of \texttt{std::vector}, which makes it a drop-in replacement for NWGraph’s adjacency structures. This allows NWGraph’s traversal algorithms, such as BFS, to be directly applied to distributed graphs with minimal code changes.
The \texttt{hpx::partitioned\_vector} exposes an API similar to that of a C++ \texttt{std::vector}, making it a straightforward drop-in for the existing NWGraph data structures. 
As such, any C++ algorithm that would work with \texttt{std::vector} readily works with \texttt{hpx::partitioned\_vector} as well.
For this reason, NWGraph's naïve sequential BFS algorithm functions when applied to an instance of a distributed graph (see Listing \ref{fig:bfs-1}).

However, BFS poses unique challenges. The size of the active frontier changes dramatically at different levels of the traversal, leading to severe load imbalance across partitions. Additionally, checking and updating visited vertices often requires fine-grained communication between localities. To address this, we implemented BFS using HPX’s asynchronous primitives. Frontier expansion is launched as a set of lightweight HPX tasks, which communicate updates asynchronously through \texttt{hpx::future} objects. This removes the need for global synchronization after every level of traversal and overlaps communication with local computation.

The BFS implementation is written in a similar style to the sequential NWGraph version, but the use of segmented iterators and asynchronous task execution enables scaling across distributed memory (see listing ~\ref{fig:bfs-2}). 

Due to asynchronous execution, the parent update must now occur atomically. In Listing~\ref{fig:bfs-2}, this is handled by the \texttt{set\_parent} function, which performs the update using \texttt{compare\_exchange} to ensure correctness (the implementation is omitted for brevity).

\subsection{Page Rank}

PageRank is a centrality algorithm originally introduced to rank web pages, but now widely applied in domains such as social networks and recommendation systems. The algorithm simulates a random walk in which, at each step, a walker either follows an outgoing edge or jumps to a uniformly chosen vertex. The score of a vertex $u$ is defined as:

\begin{equation}
\pi(u) = (1 - \alpha) \cdot \frac{1}{n} + \alpha \cdot \sum_{v \in N_{\text{in}}(u)} \frac{\pi(v)}{|N_{\text{out}}(v)|},
\end{equation}

This formulation is based on the random surfer model proposed by Brin and Page~\cite{brin1998anatomy}, where $\alpha$ is the damping factor (typically set to 0.85), $n$ is the total number of nodes, $N_{\text{in}}(u)$ represents the set of in-neighbors of $u$, and $|N_{\text{out}}(v)|$ denotes the out-degree of vertex $v$.
Distributed PageRank requires iteratively exchanging score updates across partitions. This makes it communication-intensive, since each iteration involves sending or pulling contributions from neighbors across the network. Load imbalance is another issue: high-degree vertices send a large number of updates, while low-degree vertices contribute very little. Synchronization across iterations can also limit scalability~\cite{meng2024surveydistributedgraphalgorithms}.

Our implementation follows the standard iterative formulation and is structured into three phases per iteration:

\paragraph{Contribution Accumulation.}  
For each vertex residing on the current locality, we compute its outgoing contribution as
\[
\text{contrib}[i] = \frac{\text{page\_rank}[i]}{\text{degrees}[i]}.
\]
The contributions are accumulated in a partitioned vector that stores a floating-point value per vertex.
\begin{itemize}
    \item \textbf{Local neighbors:} contributions to local neighbors are applied directly.
    \item \textbf{Remote neighbors:} for neighbors owned by a different locality, an asynchronous remote action is issued. The contribution is computed at the remote locality and sent back, atomically updating the destination vertex.
\end{itemize}

\paragraph{Rank Update.}  
After contributions are collected, each vertex updates its PageRank value according to:
\[
\text{page\_rank}[i] = \text{base\_score} + \alpha \cdot z,
\]
where $z$ is the sum of contributions from all incoming neighbors.

\paragraph{Error Computation.}  
To track convergence, the local error is computed as the sum of absolute differences between the old and new PageRank values for all vertices owned by the locality.

This distributed design leverages HPX’s asynchronous remote actions and atomic updates on partitioned data structures, enabling efficient overlap of computation and communication while preserving the sequential semantics of the algorithm.

\section{Results}

We implemented distributed versions of Breadth-First Search (BFS) and PageRank using HPX and evaluated their performance against the Boost Graph Library (BGL). Our experiments focused on understanding how well the algorithms scale with increasing numbers of processing units and how communication overhead impacts performance. Input graphs are generated using the Erd\"os-R\'enyi ("urand") model, and are of varying size (e.g., urand25 has $2^{25}$ vertices).

\begin{figure}[tpb] 
  \centering
    \includegraphics[width=0.8\columnwidth]{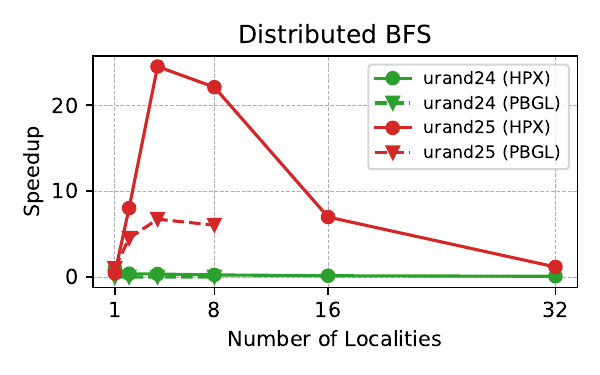}
  \caption{\footnotesize Distributed BFS runtime on GAP graphs: Boost vs our distributed implementation (HPX) }
  \label{fig:DBFS}
\end{figure}

Our first experiment was on a test machine with Intel Xeon Ice Lake processors,
on up to 32 nodes (64 cores each). Figure \ref {fig:DBFS} shows the comparison of our implementation over the Boost library. 
%
For each experiment, we measured the runtime of our distributed BFS implementation (DBFS) and compared it against Boost’s BFS implementation. The x-axis represents the number of nodes in the distributed system (ranging from 1 to 32), and the y-axis shows the corresponding speedup relative to the fastest sequential implementation for that graph size. This allows us to evaluate the scalability and performance improvements of our DBFS implementation as we increase the number of nodes. Our preliminary results indicate that the HPX-based BFS achieves better load distribution and reduced synchronization overhead compared to PBGL. 


\begin{figure}[tpb] 
  \centering
    \includegraphics[width=0.8\columnwidth]{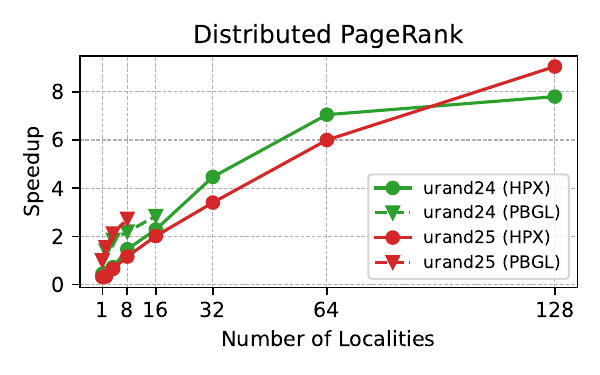}
  \caption{\footnotesize Distributed Page Rank runtime on GAP graphs: Boost vs our distributed implementation (HPX) }
  \label{fig:DPGR}
\end{figure}

In the second experiment, we applied our new PageRank algorithm prototype as well as the Boost PageRank implementation to the random graphs.
Figure \ref{fig:DPGR} shows this comparison.
In our very initial implementation of the distributed PageRank algorithm, the performance was significantly worse than the Boost library. After applying several optimizations, our prototype (HPX) has improved considerably and is now closer to Boost’s performance, although it still lags behind. We are optimistic that further optimization will allow us to close this gap and achieve even better performance; however, this remains a focus for future work.


\section{Future Work and Conclusion}

In this work, we have taken the first steps toward building a distributed graph analytics framework on top of HPX. As a proof of concept, we implemented distributed versions of Breadth-First Search (BFS) and PageRank and compared their performance to the Boost Graph Library. These initial benchmarks show that our approach is both feasible and promising. HPX’s asynchronous runtime makes it possible to overlap communication with computation, and its global address space simplifies the design of distributed data structures such as partitioned vectors. Even at this early stage, our experiments demonstrate the potential for HPX to provide a scalable and flexible alternative for graph analytics. For PageRank, however, our current implementation still lags behind Boost in terms of performance, indicating that further optimization is necessary.

Our next goal is to broaden the scope of algorithms supported. We plan to extend our implementations to cover the full set of algorithms available in NWGraph, including traversal, centrality, and pattern-matching algorithms. This will allow us to build a systematic benchmark suite for distributed graph processing in HPX and carry out detailed comparisons with established frameworks such as Boost, Gunrock, and GraphBLAS. Such an evaluation will help us understand where HPX’s execution model excels and where further runtime or algorithmic optimizations are necessary.

In addition to expanding algorithm coverage, we intend to investigate runtime adaptivity more deeply. A central vision of this research is that graph applications should run efficiently on both current and future architectures without requiring developers to rewrite their code. By integrating adaptive execution policies, partitioned data structures, and asynchronous communication, we aim to achieve performance portability together with programmer productivity. This combination positions HPX as not only a runtime for traditional HPC workloads, but also a strong platform for irregular, communication-intensive applications such as large-scale graph analytics.

The broader impact of this work goes beyond technical contributions. Graph processing plays a vital role in many domains, from bioinformatics and cybersecurity to social networks and recommendation systems. By lowering the complexity of writing and maintaining distributed graph algorithms, our research reduces the barrier for scientists and application developers who rely on these techniques but cannot maintain highly specialized parallel code. We also aim to support technology transfer, bridging the gap between academic innovation and industry adoption by producing reusable software components and runtime support that industrial partners can confidently use.

Another key aspect of our approach is the use of the adaptive executor \texttt{adaptive\_core\_chunk\_size} ~\cite{mohammadiporshokooh2025adaptively, mohammadiporshokooh2025new}, which automatically adjusts execution parameters such as the number of processing units and chunk sizes based on workload characteristics. This mechanism has already shown potential in improving performance for certain algorithms, and we plan to further leverage it in our future implementations.

Overall, these initial results highlight the potential of HPX for high-performance distributed graph processing and provide a foundation for ongoing research to develop a comprehensive, efficient, and scalable library of parallel graph algorithms.


\bibliographystyle{unsrt}
\bibliography{refs}

\end{document}